\documentclass[twocolumn,showpacs,preprintnumbers,amsmath,amssymb,superscriptaddress,pra]{revtex4}
\usepackage{graphicx}
\begin{document}
\title{Shear viscosity of  $\mathbf{p}$\,-wave superfluid Fermi gas with weak interaction at low temperatures}
\author{Soudabe Nasirimoghadam}
\address{Department of Physics, Sirjan University of Technology, Sirjan, Iran}
\author{Roohollah Aliabadi}
\address{Department of Physics, Fasa University, Fasa, Iran}
\author{Mohamad Ali Shahzamanian}
\address{Department of Physics, Faculty of Sciences, University of Isfahan, 81744, Isfahan, Iran}
\date{\today}
\begin{abstract}
The shear viscosity tensor of the superfluid Fermi gas in $p$\,-wave state with weak interaction is calculated at low temperatures, by using the Boltzmann equation approach. We consider  the transition probabilities for the binary, decay and coalescence processes. We show that only the binary processes are dominated at low temperatures and  the components of shear viscosity $\eta_{xy}$, $\eta_{xx}$, $\eta_{yy}$ are proportional to $(1/T)^{2}$, and $\eta_{xz}$, $\eta_{yz}$ and $\eta_{zz}$ are proportional to $(1/T)^{4}$ and $(1/T)^{6}$, respectively.
\end{abstract}
 \maketitle

\section{Introduction}\label{SEC1}
It is possible to create molecular Bose\,-Einstein condensates from weakly bound molecules consisting of two fermion atoms. The molecular condensates are formed when the scattering length is positive, corresponding to an effective repulsion between atoms. If the interaction is attractive, the atoms may pair up in a manner similar to the way in which electrons or atoms of liquid helium three form Cooper pairs in superconducting metals or superfluid helium three, respectively.

The ultracold atoms fermion gases such as $^{6}Li$ and $^{40}K$ undergo to the superfluid state \cite{You,Combescot,Zhang}. Salomon$^{,}$s group has reported evidence of $p$\,-wave molecules by sweeping a Fermi gas of $^{6}Li$ through a $p$ resonance \cite{Salomon}. It is therefore conceivable that the $p$\,-wave angular momentum fermion superfluids can be realized in the ultracold fermion atoms, specially in $^{6}Li$. Ho and Diener \cite{Ho} by using the generalized Nozieres and Schmitt-Rink potential \cite{Nozieres} near the scattering resonance have obtained an analytic solution of the gap equation. They showed that the ground state of $l=1$ superfluid is an orbital ferromagnet with pairing wave function $Y_{11}$.

However, the successful experimental realization of the superfluid from the BCS\,-BEC in ultracold trapped Fermi gases has an initial considerable contribution in the study of the BCS\,-BEC crossover or the regions near it. An optically trapped Fermi gas of $^{6}Li$ or $^{40}K$ is a rich system in which the strength of interaction between atoms can be controlled by applying a variable bias magnetic field tuned near a Feshbach resonance \cite{Leggett}. It is generally believed that a high temperature superfluid exists at unitarity, with a transition temperature of the order of the Fermi temperature, $T_{c}=(0.29\pm0.02)\,T_{F}$ \cite{Rupak}.

One of the most fascinating properties of ultracold trapped Fermi gases in the BCS regime, is the shear viscosity. The viscous relaxation rate can be extracted from the attenuation of collective modes \cite{Bruun1} or the frequency-dependent shear viscosity of the unitary Fermi gas can be measured experimentally using Bragg spectroscopy \cite{Taylor}.

The shear and bulk viscosities of $^{6}Li$ in the weak-coupling limit, for temperatures near $T_{c}$ and zero have been calculated by Shahzamanian and Yavari for the $s$\,-wave superfluid \cite{Shah1,Shah2}. They showed that the shear viscosity decreases as $(1-T/T_{c})$ with temperature, when passing through the transition temperature, and goes to a constant at very low temperatures. This constant is inversely proportional to the square of the scattering length. These results can be generalized to the triple superfluid state which has been considered by Modawi and Leggett \cite{Modawi}. They obtained a triple spin superfluid state with a gap function independent of the momentum direction plus a normal excitations part. The shear viscosities of superfluid part and normal part are constant and proportional to $1/T^{2}$ at low temperatures, respectively. If we ignore the interaction between these two fluids the dominated viscosity hence, is that of the normal excitations. The value of the second viscosity is temperature independent and is proportional to the inverse of the square of scattering length in the limit of low temperatures, and in the near transition temperature behaves as inverse of the gap parameter \cite{Shah2}.

Bruun and Smith by taking into account the effect of the medium on the scattering matrix for a homogeneous gas in the unitarity limit showed that the viscous and thermal relaxation rates are increased by nearly an order of magnitude compared to their values in the absence of medium effect due to the Cooper instability at $T_{c}$. The viscosity of a homogeneous system in the normal and superfluid gases of fermions is calculated analytically at low and high temperature limits for the normal and $s$\,-wave superfluid  \cite{Bruun1,Bruun2}.

The shear viscosity of the unitarity ultracold gases, similar to the bosonic superfluid $^{4}He$ (which has been considered by Wilks \cite{Wilks}), has been calculated by Rupak and Sch\"{a}fer \cite{Rupak}. The latter used the Boltzmann equation approach and showed that only the binary $2\!\!\rightarrow\!\!2$ collisions in which the number of particles is conserved, are dominated in the collision integrals at low temperatures, whereas the contribution of the splitting processes in which the number of particles is not conserved can be neglected. The scattering amplitude in the binary collision process has been calculated by many body approach. Finally they could obtain a limit on the shear viscosity by variational method. The shear viscosity of bosonic superfluid at low temperatures, scales as ${\xi^{5}}/{T^{5}}$, where the universal parameter $\xi$ relates chemical potential and the Fermi energy, $\mu=\xi\varepsilon_{F}$ \cite{Rupak}.

The Kubo formulas approach has been used to calculate the shear and bulk viscosities of strongly interacting ultracold Fermi, Bose gases by Taylor and Randeria \cite{Taylor}, Shahzamanian and Yavari \cite{Shah2}. The former derived sum rules for the shear and bulk viscosity spectral functions, $\eta(w)$ and $\zeta(w)$, respectively, of any Bose or Fermi system with arbitrary two\,-body interaction, and also at unitarity, the bulk or second viscosity spectral function vanishes at all frequencies and all temperatures. The latter by taking into account the contributions of the interactions between condensate and noncondensate atoms and between noncondensate atoms could obtain the viscous relaxation times and calculated the value of shear viscosity of a trapped Bose\,-condensed atomic gas in different limits of temperatures. The shear and bulk viscosity in a BCS-BEC crossover scheme have been computed by Levin group \cite{Levin}. They extended the BCS-Leggett ground state to nonzero temperature and obtained two contributions to the square of pairing gap, corresponding to condensed and to non-condensed pairs. Their results show that the shear viscosity goes to zero at $T\longrightarrow0$.

In this paper, we consider the shear viscosity of $p$\,-wave superfluid Fermi gas, in the state $Y_{11}$ at low temperatures by using a potential model that is a generalization of the separable potential used by Nozieres and Schmitt-Rink \cite{Nozieres}. Nozieres and Schmitt-Rink studied a gas of fermions interacting via an attractive potential. They introduced the separable potential for $s$\,-wave fermion gases. Ho and Diener generalized it to the $p$\,-wave fermion superfluid gases \cite{Ho}. Here we consider only the shear viscosity components of superfluid. As we mentioned, the shear viscosity of Bosic part, has been calculated by Rupak and Sch\"{a}fer \cite{Rupak}. In this paper we suppose that these superfluids are mixture independently.

By using  the Boltzmann equation approach and considering effective interaction between quasiparticles in the superfluid state, we obtain the transition probabilities of possible processes at low temperatures. Since in a superfluid the quasiparticles number is not conserved, then other processes can occur, such as decay processes in which one quasiparticle decays into three and coalescence processes in which three quasiparticles coalesce to produce one, whereas in a normal Fermi gas at low temperatures, only the binary processes are dominated.

Transition probabilities are obtained according to a generalization of the potential model used by Nozieres and Schmitt-Rink. More clearly, this potential model is independent of angles between the momenta of quasiparticles. We calculate the shear viscosity components of superfluid Fermi gas at low temperatures by Sykes and Brooker procedure \cite{Sykes}.

\section{Transition probabilities}\label{2}
The effective interaction between quasiparticles in the superfluid states, is obtained by performing Bogoliuobov transformation on the normal interaction.

The potential energy in BCS Hamiltonian is \cite{Tilley}
 \begin{align}
V=\sum_{\vec{p}_{1},\vec{p}_{2},\vec{p}_{3},\vec{p}_{4}}V_{\vec{q}}^{}\, a_{\vec{p}_{3}}^{\dag}\,a_{\vec{p}_{4}}^{\dag}\,a_{\vec{p}_{2}}^{}\,a_{\vec{p}_{1}}^{},\label {Potential1}
\end{align}
where $\vec{q}=\vec{p}_{3}-\vec{p}_{1}=\vec{p}_{4}-\vec{p}_{2}$. The Bogoliubov transformation, between the normal quasiparticle creation $a_{\vec{p},\sigma}^{\dag}$ and annihilation $a_{\vec{p},\sigma}$ operators, and creation and  annihilation  operators $\alpha_{\vec{p},\sigma}^{\dag}$ and $\alpha_{\vec{p},\sigma}$ in the superfluid may be written as
\begin{align}
a_{\vec{p},\sigma}=u_{\vec{p}}\,\alpha_{\vec{p},\sigma}
+\sigma\, v_{\vec{p}}^{}\, \alpha_{-\vec{p},-\sigma}^{\dag}.\label{Operator1}
\end{align}
For the superfluid in $p$\,-wave state, we have the following properties between $u_{\vec{p}}$ and $v_{\vec{p}}$ \cite{Takagi}
\begin{align}
u_{-\vec{p}}=u_{\vec{p}}\qquad;\qquad    v_{-\vec{p}}=-v_{\vec{p}}.\label{Bogo.coe}
\end{align}
By using Eqs. \!\!\eqref{Operator1} and \eqref{Bogo.coe} in Eq. \!\!\eqref{Potential1} and the anticommutation relations for the quasiparticle creation and annihilation operators, the potential energy is obtained \cite{Shah3}
\begin{align}
V=\sum_{\vec{p}_{1},\vec{p}_{2},\vec{p}_{3},\vec{p}_{4}}
&V_{\vec{q}}[(u_{\vec{p}_{3}}\alpha_{\vec{p}_{3},\sigma}^{\dag}+\sigma v_{\vec{p}_{3}}^{} \alpha_{-\vec{p}_{3},-\sigma})\nonumber\\
\times&(u_{\vec{p}_{4}}\alpha_{\vec{p}_{4},\sigma'}^{\dag}+\sigma' v_{\vec{p}_{4}}^{} \alpha_{-\vec{p}_{4},-\sigma'})\nonumber\\
\times&(u_{\vec{p}_{2}}\alpha_{\vec{p}_{2},\sigma'}+\sigma' v_{\vec{p}_{2}}^{} \alpha_{-\vec{p}_{2},-\sigma'}^{\dag})\nonumber\\
\times&(u_{\vec{p}_{1}}\alpha_{\vec{p}_{1},\sigma}+\sigma v_{\vec{p}_{1}}^{} \alpha_{-\vec{p}_{1},-\sigma}^{\dag})],\label{Potential2}
\end{align}
which contains the terms\,$\alpha_{\vec{p}_{3},\sigma}^{\dag}\alpha_{\vec{p}_{4},-\sigma'}^{\dag}\alpha_{-\vec{p}_{2},\sigma'}^{\dag}\alpha_{\vec{p}_{1}
,\sigma}^{}$,\\$\alpha_{\vec{p}_{3},\sigma}^{\dag}\alpha_{\vec{p}_{1},\sigma}^{}\alpha_{-\vec{p}_{4},-\sigma'}^{}\alpha_{\vec{p}_{2},\sigma'}^{}$
, $\alpha_{\vec{p}_{3},\sigma}^{\dag}\alpha_{\vec{p}_{4},\sigma'}^{\dag}\alpha_{\vec{p}_{2},\sigma'}^{}\alpha_{\vec{p}_{1},-\sigma}^{}$
,\\$\alpha_{\vec{p}_{3},\sigma}^{\dag}\alpha_{\vec{p}_{4},\sigma'}^{\dag}\alpha_{-\vec{p}_{2},-\sigma'}^{\dag}\alpha_{-\vec{p}_{1},-\sigma}^{\dag}$and $\alpha_{-\vec{p}_{3},-\sigma}^{}\alpha_{-\vec{p}_{4},-\sigma'}^{}\alpha_{\vec{p}_{2},\sigma'}^{}\\\alpha_{\vec{p}_{1},\sigma}^{}$. These terms convert a quasiparticle into three, coalescence three quasiparticles into one, convert two quasiparticles into two, create four quasiparticles from the condensate and scatter four quasiparticles into condensate, respectively. The last two processes are not allowed, because in each process the total energy should be conserved.

The  transition probability, for example, the decay process is given by
\begin{align}
W_{13}=|\langle...;\vec{p}_{3},\sigma;\vec{p}_{4},\sigma';-\vec{p}_{2},-\sigma';...|V|...;\vec{p}_{1},\sigma;...\rangle|^{2},\label{Transition probability}
\end{align}
when Eq. \!\!\!\!\!(\ref{Potential2}) is substituted in Eq. \!\!\!\!\!(\ref{Transition probability}) only the term $\alpha_{\vec{p}_{3},\sigma}^{\dag}\alpha_{\vec{p}_{4},-\sigma'}^{\dag}\alpha_{-\vec{p}_{2},\sigma'}^{\dag}\alpha_{\vec{p}_{1}
,\sigma}$ will be entered the calculations ,thus the transition probability may be written as
\begin{align}
W&_{13}(\vec{p}_{1},\vec{p}_{2},\vec{p}_{3},\vec{p}_{4};\sigma,\sigma')\nonumber\\
&=\frac{1}{4}|(u_{\vec{p}_{4}}u_{\vec{p}_{3}}u_{\vec{p}_{1}}v_{\vec{p}_{2}}+
v_{\vec{p}_{4}}v_{\vec{p}_{3}}v_{\vec{p}_{1}}u_{\vec{p}_{2}})\nonumber\\
&\times[\sigma'(V_{\vec{p}_{3}-\vec{p}_{1}}+V_{\vec{p}_{1}-\vec{p}_{3}})-\sigma'\delta_{\sigma,\sigma'}(V_{\vec{p}_{4}-\vec{p}_{1}}+V_{\vec{p}_{1}-\vec{p}_{4}})]\nonumber\\
&+(u_{\vec{p}_{3}}u_{\vec{p}_{2}}u_{\vec{p}_{1}}v_{\vec{p}_{4}}+v_{\vec{p}_{3}}v_{\vec{p}_{2}}v_{\vec{p}_{1}}u_{\vec{p}_{4}})\nonumber\\
&\times[-\sigma'(V_{\vec{p}_{3}-\vec{p}_{1}}+V_{\vec{p}_{1}-\vec{p}_{3}})+\sigma'\delta_{\sigma,-\sigma'}(V_{-\vec{p}_{1}-\vec{p}_{2}}
+V_{\vec{p}_{1}+\vec{p}_{2}})]\nonumber\\
&+(u_{\vec{p}_{4}}u_{\vec{p}_{2}}u_{\vec{p}_{1}}v_{\vec{p}_{3}}+v_{\vec{p}_{4}}v_{\vec{p}_{2}}v_{\vec{p}_{1}}u_{\vec{p}_{3}})\nonumber\\
&\times[\sigma'\delta_{\sigma,\sigma'}(V_{\vec{p}_{4}-\vec{p}_{1}}+V_{\vec{p}_{1}-\vec{p}_{4}})\nonumber\\
&-\sigma'\delta_{\sigma,-\sigma'}(V_{-\vec{p}_{1}-\vec{p}_{2}}+V_{\vec{p}_{1}+\vec{p}_{2}})]|^{2}.\label{w}
\end{align}

In general, as is seen from Eq. \!(\ref{w}) $W_{13}$ is, a function of all the momenta through potentials, spins and temperature.
The transition probabilities of other processes such as, $W_{31}(\sigma,\sigma')$ and $W_{22}(\sigma,\sigma')$ are written in terms of $u$ and $v$, the results of these calculations are in the Appendix.

The Bogoliubov coefficients $u_{\vec{p}}$ and $v_{\vec{p}}$ can be written as
\begin{align}
u_{\vec{p}}^{2}=\frac{1}{2}(1+\frac{\varepsilon_{\vec{p}}}{E_{\vec{p}}})\qquad;\qquad v_{\vec{p}}^{2}=\frac{1}{2}(1-\frac{\varepsilon_{\vec{p}}}{E_{\vec{p}}}),\label{u,v}
\end{align}
where $E_{\vec{p}}^{2}=\varepsilon_{\vec{p}}^{2}+\Delta_{\vec{p}}^{2}$, $\varepsilon_{\vec{p}}$ is the normal state quasiparticle energy measured with respect to the chemical potential and $\Delta_{\vec{p}}$ is the magnitude of the gap in direction $\vec{p}$ on the Fermi surface \cite{Takagi}.
$\Delta_{\vec{p}}=\Delta(T)\sin\theta_{p}$, where $\Delta(T)$ is the maximum gap and $\theta_{p}$ is the angle between the quasiparticle momentum and gap axis $\hat{l}$ that is supposed to be in the direction of $z$ axis.\\

At low temperatures, because more quasiparticles are gathered in the nodes of gap, thus we have $\sin\theta_{p_{i}}\simeq0$ and $\Delta_{p}\simeq0$ and $E_{\vec{p}}\simeq\varepsilon_{\vec{p}}$. In this temperature region, Bogoliubov coefficients can be approximated as $u_{\vec{p}}\simeq1$ and $v_{\vec{p}}\simeq0$.
By using these approximations, we find the transition probabilities as
\begin{align}
&W_{13}(\uparrow\downarrow)\simeq0\quad,\quad W_{13}(\upuparrows)\simeq0,\nonumber\\
&W_{31}(\upuparrows)\simeq0\quad,\quad W_{31}(\uparrow\downarrow)\simeq0,
\end{align}
\begin{align}
W_{22}(\uparrow\downarrow)=\frac{1}{4}|(V_{\vec{p}_{1}-\vec{p}_{3}}+V_{\vec{p}_{3}-\vec{p}_{1}})|^{2},
\end{align}
\begin{align}
W_{22}(\upuparrows)=\frac{1}{4}|(V_{\vec{p}_{1}-\vec{p}_{3}}+V_{\vec{p}_{3}-\vec{p}_{1}})-(V_{\vec{p}_{3}-\vec{p}_{2}}+V_{\vec{p}_{2}-\vec{p}_{3}})|^{2},
\end{align}
above equations indicate that the transition probabilities are functions of the potential. In this paper, as we mentioned we are using the model of potential, that is a generalization of the separable potential used by Nozieres and Schmitt-Rink, and can be written as \cite{Ho}
\begin{align}
V_{l}(p,p')=\lambda_{l}W_{l}(p)W_{l}(p')\quad,\quad W_{l}(p)=\displaystyle{\frac{(\frac{p}{p_{0}})^l}{[1+(\frac{p}{p_{0}})^2]^{\frac{l+1}{2}}}},
\end{align}
where, $p_{0}^{-1}$ is the range of potential (in this paper we put $\hbar\equiv k_{B}\equiv1$).
This potential for $p$\,-wave state may be written as
\begin{align}
V_{1}(p_{1},p_{3})=\lambda_{1}\frac{p_{0}^{2}p_{1}p_{3}}{(p_{0}^{2}+p_{1}^{2})(p_{0}^{2}+p_{3}^{2})}.
\end{align}
Because quasiparticles are gathered near the Fermi surface at low temperatures, therefore their wave numbers are nearly Fermi wave number. Thus, one may write
\begin{align}
V_{1}(p_{1},p_{3})=V_{1}(p_{2},p_{3})\simeq\lambda_{1}\frac{p_{0}^{2}p_{F}^{2}}{(p_{0}^{2}+p_{F}^{2})^{2}}.
\end{align}
For estimating this potential, we should know the order of magnitude of $p_{F}$ and $p_{0}$ to each other.
The kinetic energy of superfluid is ${p_{0}^{2}}/{2m}-\varepsilon_{F}$. From equating the potential energy with the kinetic energy, one may obtain $p_{0}$ as following
\begin{align}
p_{0}^{2}\simeq p_{F}^{2}+p_{\lambda}^{2},
\end{align}
where, $p_{\lambda}=(m\lambda_{1}/2)^{\frac{1}{2}}$. This results $p_{0}>p_{F}$. With this assumption, the potential energy may be written as
\begin{align}
V_{1}(p_{1},p_{3})\simeq\frac{\lambda_{1}}{4},
\end{align}
with applying this value of potential, the transition probabilities are as follows
\begin{align}
W_{22}(\uparrow\downarrow)=|V_{1}(p_{1},p_{3})|^{2}\simeq\frac{\lambda_{1}^{2}}{16},
\end{align}
\begin{align}
W_{22}(\upuparrows)=|V_{1}(p_{1},p_{3})-V_{1}(p_{2},p_{3})|^{2}\simeq0.
\end{align}

It is noted that only the binary processes are dominated in the limit of low temperatures and the transition probabilities of other processes have nearly zero value. Here, we emphasize that only the spin up and spin down quasiparticles in the binary processes are contributed to the transition probabilities at low temperatures which is happened in superfluid helium three with strong interaction, too \cite{Shah3}. The binary processes are dominated also in the BEC regime at low temperatures \cite{Rupak}.
\section{Collision integral}\label{3}
The kinetic equation for the quasiparticle distribution function of superfluid, $\nu_{p,\sigma}$, can be written as \cite{Volhardt}
\begin{align}
\frac{\partial\nu_{p,\sigma}}{\partial t}+\frac{\partial\nu_{p,\sigma}}{\partial \vec{r}}\frac{\partial E_{p,\sigma}}{\partial \vec{p}}
-\frac{\partial\nu_{p,\sigma}}{\partial \vec{p}}\frac{\partial E_{p,\sigma}}{\partial \vec{r}}=I(\nu_{p,\sigma}),\label{Collision integral}
\end{align}
where, $I(\nu_{p,\sigma})$ is the collision integral. Let us assume a disturbance of the form
\begin{align}
\nu_{\sigma}(p,r)=\nu^{0}(p)+\delta\nu_{\sigma}(p),\label{Distribution function}
\end{align}
where, in general both $\nu(p)$ and $\delta\nu_{\sigma}(p)$ are functions of the variable $r$, and $\nu^{0}(p)$ is the distribution function of equilibrium superfluid state. We define the function $\psi_{\sigma}(p)$ by
\begin{align}
\delta\nu_{\sigma}(p)=-\frac{1}{T}\nu^{0}(p)(1-\nu^{0}(p))\psi_{\sigma}(p),\label{Nu}
\end{align}
where, $\nu^{0}(p)=[\exp(E_{p}^{0}-\vec{p}\cdot\vec{u})/T+1]^{-1}$ and $u$ is the slightly inhomogeneous velocity in the gas. Substituting Eq. \!\!(\ref{Distribution function}) in Eq. \!\!\!(\ref{Collision integral}), keeping the terms which contribute to the shear viscosity, and supposing $\vec{u}$ and $\vec{\nabla}\cdot\vec{u}$ are zero at the point considered, to first order in $\delta\nu_{\sigma}(p)$ we have
\begin{align}
-\frac{1}{2}\frac{\partial\nu^{0}}{\partial E_{p,\sigma}}p_{i}\frac{\partial E_{p,\sigma}}{\partial p_{k}}
(\frac{\partial u_{i}}{\partial r_{k}}+\frac{\partial u_{k}}{\partial r_{i}}-\frac{2}{3}\delta_{ij}\vec{\nabla}\cdot\vec{u})
=I(\delta\nu_{\sigma}(p)).\label{12}
\end{align}

Now, by using Eq. \!(\ref{Distribution function}) and keeping the terms to first order in $\psi_{\sigma}(P)$, the linearized collision terms in the Boltzmann equation can be written as
\begin{align}
I_{22}&=\frac{(m^{*})^{3}T}{64\pi^{5}}\int W_{22}\frac{sin\theta}{cos(\frac{\theta}{2})}
d\theta d\varphi d\varphi_{2}\int_{-\infty}^{\infty}dxdy
(\psi(t)\nonumber\\
&+\psi(x+y-t)-\psi(x)-\psi(y))\nu^{0}(t)\nu^{0}(x+y-t)\nonumber\\
&\times(1-\nu^{0}(x))(1-\nu^{0}(y)),
\end{align}
where, $\varphi_{2}$ is the azimuthal angle of $\vec{p}_{2}$ relative to $\vec{p}_{1}$, $t\equiv E_{1}/T$, $x\equiv E_{3}/T$, and $y\equiv E_{4}/T$.\\
It is noted that only two quasiparticles processes take part to the collision integral at low temperatures and only $W_{22}(\uparrow\downarrow)$ is not negligible for the Nozieres and Schmitt-Rink potential.

To solve the linearized Boltzmann equation, it is suitable to define $q(T)$ as
\begin{align}
\psi(t)& =p_{i}\frac{\partial E_{1}}{\partial p_{1k}} q(t) [\frac{\partial u_{i}}{\partial r_{k}} +\frac{\partial u_{k}}{\partial r_{i}}-\frac{2}{3}\delta_{ij}\vec{\nabla}\cdot\vec{u}].\label{Psi}
\end{align}
By expressing the bracket in Eq. \!\!(\ref{Psi}) in terms of a series of spherical harmonics, i.e.,$\sum_{m=-2}^{m=2}U_{m}p_{2}^{|m|}(cos\Theta)e^{im\Phi}$ \cite{Sykes} and substituting Eq. \!\!\!\!(\ref{Psi}) in the collision integral, the integration on $\varphi_{2}$ can be done conveniently.\\ By noting some symmetries with respect to variables in the collision integral \cite{Sykes},and performing integration on $y$, we get
\begin{align}
I_{22}& =\frac{(m^{*})^{3}T^{2}}{32\pi^{4}}\sum_{m=-2}^{m=2}U_{m}p_{2}^{|m|}(cos\Theta)e^{im\Phi}(-\frac{\partial \nu^{0}}{\partial E_{1}})\nonumber\\
&\times \int dxK(t,x)\int\frac{sin\theta d\theta d\varphi}{cos(\frac{\theta}{2})}W_{22}[q(t)+\nonumber\\
&q(-x)p_{2}(cos\theta)-q(x)({p_{2}(cos\theta_{13})+p_{2}(cos\theta_{14})})],\label{Final integral}
\end{align}
where $K(t,x)$ is
\begin{align}
K(t,x)=\frac{e^{-t}+1}{e^{-x}+1}\frac{x-t}{e^{(x-t)}-1}.
\end{align}
By substituting Eq. \!\!\!\!\!(\ref{Final integral}) in Eq. \!\!\!\!\!(\ref{12}) and considering $K(t,x)=K(-t,-x)$, we have \cite{Sykes}
\begin{align}
\int dxK(t,x)[q_{s_{\sigma}}(t)-\lambda_{2s_{\sigma}}q_{s_{\sigma}}(x)]=B_{\sigma},\label{B}
\end{align}
\begin{align}
\int dxK(t,x)[q_{a_{\sigma}}(t)-\lambda_{2s_{\sigma}}q_{a_{\sigma}}(x)]=O(TB_{\sigma}),\label{O}
\end{align}
where $q_{s_{\sigma}}(t)$ and $q_{a_{\sigma}}(t)$ are symmetric and antisymmetric parts of $q(t)$, respectively, and $\sigma$ is simply the spin index.
At low temperatures, Eq. \!(\ref{O}) is negligible, and Eq. \!(\ref{B}) is dominated \cite{Sykes}. In Eq. \!(\ref{B}), $\lambda_{2s_{\sigma}}$ and $B_{\sigma}$ are introduced as
\begin{align}
\lambda_{2s_{\sigma}}=\frac{\displaystyle {\int\frac{sin\theta d\theta d\varphi}{cos(\frac{\theta}{2})}W_{22}
[1-\frac{3}{4}(1-cos\theta)^{2}sin^{2}\varphi]}}{\displaystyle {\int\frac{sin\theta d\theta d\varphi}{cos(\frac{\theta}{2})}W_{22}}},\label{landa1}
\end{align}
\begin{align}
B_{\sigma}=\frac{16\pi^{5}}{m^{*3}T^{2}}[\int\frac{sin\theta d\theta d\varphi}{cos(\frac{\theta}{2})}W_{22}]^{-1},\label{B2}
\end{align}
where, $W_{22}$ is the transition probability for binary processes and is independent of angles. Generally, for $\sigma=\uparrow$ and $\downarrow$, $W_{22}$ stands for $W_{22}(\uparrow\downarrow)$ and $W_{22}(\downarrow\uparrow)$, respectively. \,Because in the absence of magnetic field, $W_{22}(\uparrow\downarrow)=W_{22}(\downarrow\uparrow)$, the function $W_{22}$ is the same. Therefore $W_{22}$ in Eqs. \!\!(\ref{landa1}) and (\ref{B2}) is ${\lambda_{1}^{2}}/16$ for both cases.

For obtaining $\lambda_{2s_{\sigma}}$ and $B_{\sigma}$, we note that $\theta$ is small for the neutral superfluid case and its maximum value is $\theta_{m}={\pi T}/\Delta(0)$ \cite{Shah4}, where maximum gap, $\Delta(0)$, for fermion gas is equal to $1.76T_{c}$ at low temperatures \cite{Pethick}. Finally, we obtain
\begin{align}
\lambda_{2s_{\uparrow}}=\lambda_{2s_{\downarrow}}\simeq 1-\frac{1}{32}\theta_{m}^{4},\label{landa}
\end{align}
\begin{align}
B_{2s_{\uparrow}}=B_{2s_{\downarrow}}\simeq \frac{16\pi^{5}}{m^{*3}T^{2}}\frac{1}{W_{22}\pi \theta_{m}^{2}}.\label{B3}
\end{align}
Following the Sykes and Brooker procedure \cite{Sykes}, from Eq. \!(\ref{B}) we have
\begin{align}
\int_{-\infty}^{\infty}dt\frac{d\nu^{0}}{dt}{q_{2s_{\sigma}}(t)=\frac{-2B}{\pi^{2}(1-\lambda_{2s_{\sigma}})}c(\lambda_{2s_{\sigma}})},\label{C}
\end{align}
where, $c(\lambda)$ is \cite{Sykes}
\begin{align}
c&(\lambda_{2s_{\sigma}})=\nonumber\\
&\frac{(1-\lambda_{2s_{\sigma}})}{4}\displaystyle{\sum_{n=0}^{\infty}}\quad\frac{(4n+3)}{(n+1)(2n+1)[(n+1)(2n+1)-\lambda_{2s_{\sigma}}]}\nonumber\\
& =\frac{\lambda_{2s_{\sigma}}-1}{2\lambda_{2s_{\sigma}}}[\gamma+\ln2+\frac{1}{2}\psi_{d}(s_{1})+\frac{1}{2}\psi_{d}(s_{2})],
\end{align}
where $\gamma=0.577...$ is Euler$^{,}$s constant and $\psi_{d}$ is a digamma function that is given by
\begin{align}
\psi(z+1)=-\gamma+\!\displaystyle{\sum_{n=1}^\infty}\! \frac{z}{n(n+z)},
\end{align}
and
\begin{align}
s_{1}\equiv\frac{3}{4}+\frac{1}{4}\sqrt{8\lambda_{2s_{\sigma}}+1}\quad,\quad s_{2}\equiv\frac{3}{4}-\frac{1}{4}\sqrt{8\lambda_{2s_{\sigma}}+1},
\end{align}
by calculating the expansion of series in digamma function and ignoring the small terms in those, we obtain $c(\lambda_{2s_{\uparrow}})=c(\lambda_{2s_{\downarrow}})\simeq0.75$. The value of $c(\lambda_{2s_{\sigma}})$ has been calculated in \cite{Shah3} for superfluid liquid $^{3}He$ by Pfitzner procedure \cite{Pfitzner} and is $0.77$ which is near to our obtained value $0.75$. The temperature dependence of $\lambda$ and $B$ is completely different from the $\lambda$ and $B$ that have been obtained by Shahzamanian and Afzali \cite{Shah3}.

Now, we will proceed to calculate the shear viscosity in the next section by using the calculated values of $B_{\sigma}$, $\lambda_{2s_{\sigma}}$ and
$c(\lambda_{2s_{\sigma}})$.
\section{Shear viscosity}\label{4}
The shear viscosity is a fourth-rank tensor, which is defined by the relation
\begin{align}
\pi_{lm}=-\sum_{ik}\eta_{lmik}(\frac{\partial u_{i}}{\partial r_{k}}+\frac{\partial u_{k}}{\partial r_{i}}-\frac{2}{3}\delta_{ij}\vec{\nabla}\cdot\vec{u}),\label{Pi1}
\end{align}
where $\pi_{lm}$, the momentum flux tensor is
\begin{align}
\pi_{lm}=\int p_{l}\frac{\partial E}{\partial p_{m}}\delta\nu_{\sigma}(p)d\tau_{p},\label{Pi2}
\end{align}
when Eq. \!\!\!\eqref{Psi} is substituted in Eq. \!\!\!\eqref{Nu} and then is inserted in Eq. \!\!(\ref{Pi2}) and is compared with Eq. \!\!(\ref{Pi1}), we have
\begin{align}
\eta_{lmik}=&-\frac{4p_{F}^{5}}{(2\pi)^{3}m^{*}}\int d\Omega_{p}\hat{p}_{l}\hat{p}_{m}\hat{p}_{i}\hat{p}_{k}\nonumber\\
&\times[ \int dt \frac{\partial \nu^{0}}
{\partial t}q_{2s_{\uparrow}}(t)+\int dt \frac{\partial \nu^{0}}
{\partial t}q_{2s_{\downarrow}}].\label{eta}
\end{align}
Using Eq. \!(\ref{C}) in Eq. \!(\ref{eta}), we get
\begin{align}
\eta_{lmik}=&\frac{p_{F}^{5}}{\pi^{5}m^{*}}\int d\Omega_{p}\hat{p}_{l}\hat{p}_{m}\hat{p}_{i}\hat{p}_{k}\nonumber\\
&(\frac{B_{\uparrow}}{1-\lambda_{2s_{\uparrow}}}c(\lambda_{2s_{\uparrow}})
+\frac{B_{\downarrow}}{1-\lambda_{2s_{\downarrow}}}c(\lambda_{2s_{\downarrow}}) ),
\end{align}
because $B_{\uparrow}=B_{\downarrow}$, $\lambda_{2s_{\uparrow}}=\lambda_{2s_{\downarrow}}$ and $c(\lambda_{2s_{\uparrow}})=c(\lambda_{2s_{\downarrow}})$ and also the $p$\,-wave superfluid has two nodes, we may write
\begin{align}
\eta_{lmik}=\frac{4p_{F}^{5}}{\pi^{5}m^{*}}\int d\Omega_{p}\hat{p}_{l}\hat{p}_{m}\hat{p}_{i}\hat{p}_{k}
(\frac{B}{1-\lambda_{2s}}c(\lambda_{2s})).
\end{align}
Substituting the values of $B_{\sigma}$, $\lambda_{2s_{\sigma}}$ and $c(\lambda_{2s_{\sigma}})$ and taking the angular integrations, we obtain
\begin{align}
\eta_{zz}=151.30\frac{p_{F}^{5}}{m^{*4}W_{22}}\frac{T_{c}^{4}}{T^{6}},\label{X}
\end{align}
\begin{align}
\eta_{xz}=\eta_{yz}=120.52\frac{p_{F}^{5}}{m^{*4}W_{22}}\frac{T_{c}^{2}}{T^{4}},\label{Y}
\end{align}
\begin{align}
\eta_{xx}=\eta_{yy}=3\eta_{xy}=192\frac{p_{F}^{5}}{m^{*4}W_{22}}\frac{1}{T^{2}}.\label{Z}
\end{align}
These viscosity components obtained in Eqs. \!\!(\ref{X})$-$(\ref{Z}). The transition probability $W_{22}$ is written as ${\lambda_{1}^{2}}/16$, where $\lambda_{1}$ is the strength of interaction. $p_{F}$ is given by $1.91\,({N^{1/6}}/\bar{a})$, where $N$ is the number of quasiparticles and $\bar{a}$ is the harmonic oscillator characteristic length which is written as $\bar{a}=({1}/{m\bar{\omega}})^{1/2}$ \cite{Pethick}, where, $m$ and $\bar{\omega}$ are atomic mass and angular frequency, respectively. If we use the values of $\bar{\omega}$ from \cite{Pethick} and $m$ from \cite{Shimi}, we get the harmonic oscillator lengths for $^{6}Li$ and $^{40}K$, equal to $0.30\times10^{-5}$, $0.13\times10^{-5}$, respectively. Then if we take $N=10^{6}$ \cite{Pethick}, hence the values of $p_{F}$ for $^{6}Li$ and $^{40}K$ are $6.72\times10^{-28}$ and $15.50\times10^{-28}$, respectively (these calculations are in SI units).

Now we write the shear viscosity components in Eqs. \!(\ref{X})$-$(\ref{Z}) in terms of the shear viscosity at $T_{c}$, $\eta(T_{c})$. At the neutral Fermi gas, $\eta(T_{c})$ may be written as \cite{Pines}
\begin{align}
\eta(T_{c})=\frac{16}{45T_{C}^{2}}\frac{m^{*}\hbar^{3}v_{F}^{5}}{\displaystyle \langle\frac{W_{22}sin^{4}(\frac{\theta}{2})sin^{2}\varphi}{cos(\frac{\theta}{2})}\rangle},
\end{align}
where in the normal state, $W_{22}=W_{22}(\uparrow\downarrow)+\frac{1}{2}W_{22}(\uparrow\uparrow)$, when $W_{22}$ is independent of the angles, one may write
\begin{align}
\eta(T_{c})=\frac{1}{6\pi}\frac{p_{F}^{5}}{m^{*4}W_{22}T_{c}^{2}},
\end{align}
with $W_{22}\simeq({\lambda_{1}^{2}}/16)$.
 After calculating $\eta(T_{c})$, the shear viscosity components now are written as
\begin{align}
\eta_{zz}=2851.96\: \eta (T_{C})\:(\frac{T_{C}}{T})^{6},\label{etazz}
\end{align}
\begin{align}
\eta_{xz}=\eta_{yz}=2271.74\: \eta (T_{C})\:(\frac{T_{C}}{T})^{4},\label{etaxz}
\end{align}
\begin{align}
\eta_{xx}=\eta_{yy}=3\eta_{xy}=3619.14\: \eta (T_{C})\:(\frac{T_{C}}{T})^{2}.\label{etaxx}
\end{align}
It is noted that the quantities of $B$ and $\lambda$ in Eqs.\!\! (\ref{landa}) and (\ref{B3}) give the most contribution to the temperature dependence of the shear viscosity components in Eqs. \!\!(\ref{etazz})$-$(\ref{etaxx}).
\section{Conclusions}\label{5}
By using the interaction term in BCS Hamiltonian, we calculate the transition probabilities for the binary, decay and coalescence processes.
Then we obtain, the transition probabilities in terms of the generalization of the separable potential used by Nozieres and Schmitt-Rink \cite{Nozieres}.
Then we calculate the components of the shear viscosity tensor of Fermi superfluid in $p$\,-wave state. The shear viscosity tensor for a system with uniaxial symmetry, which is under our consideration, can be written in terms of the components of the symmetry axis, $\hat{l}$, with five coefficients \cite{Shah5}. It is shown that only three of the coefficients are independent.

Roobol et al.$^{,}$s results on the shear viscosity of the $A_{1}$-phase of superfluid $^{3}He$ in which only a single spin population is paired, indicate $\eta\propto(T)^{-2}$ at low temperatures \cite{Roobol}. They illustrate that in measuring the values of the viscosity, $\eta$, the values of the component $\eta_{zz}$ have more contribution than the values of the component $\eta_{xz}$. Shahzamanian and Afzali \cite{Shah3} by using the procedures of Pfitzner, and Sykes and Brooker could show these temperature dependencies in superfluid $^{3}He$-$A_{1}$.
As it is obvious the $A_{1}$-phase of superfluid $^{3}He$ contains the superfluid with spin up Cooper pairs and the normal fluid with spin down. The interaction between these fluids plays the important role in temperature dependence of the shear viscosity components.
In this paper, we obtain $\eta_{xy}$, $\eta_{xz}$ and $\eta_{zz}$ with temperature dependences as $1/{T^{2}}$, $1/{T^{4}}$ and $1/{T^{6}}$, respectively (see Eqs. \!(\ref{X}$-$\ref{Z})).
Thermal relaxation rates may be determined directly experimentally, and viscous relaxation rates can be extracted from the attenuation of collective modes \cite{Bruun1}. Our results on the viscosity components indicate that all collective modes in ultracold fermion gases in $p$\,-wave state are strongly damped.

The experimental measurements of the attenuation of collective modes in ultracold fermion gases can determine whether the superfluid state is $s$\,-wave or $p$\,-wave, since in $s$\,-wave state the shear viscosity is constant at low temperatures \cite{Bruun1,Shah1}, whereas in $p$\,-wave state all components are proportional inversely to temperature with powers of $2,4$ and $6$.

The viscosity of Bosonic part can be compared with those of Fermionic parts. The component $\eta_{zz}$ is proportional to $T^{-6}$ and will be dominated in the ultracold gases at low temperatures. The experimental results in this case are a clue to determine the state of the system.

\section*{Appendix}
In this Appendix we obtain the transition probabilities for other processes in $p$\,-wave superfluid Fermi gas.
By using Eqs. \!\!\!\eqref{Operator1} and \eqref{Bogo.coe} in Eq. \!\!\eqref{Potential1} and the anticommutation relations for the quasiparticle creation and annihilation operators, the potential energy may be written as
\begin{align*}
V=\frac{1}{2\Omega}&\sum V_{13}(\vec{p}_{1},...,\vec{p}_{4};\sigma,\sigma')\alpha_{\vec{p}_{3},\sigma}^{\dag}\alpha_{\vec{p}_{4},-\sigma'}^{\dag}\alpha_{-\vec{p}_{2},\sigma'}^{\dag}\alpha_{\vec{p}_{1},\sigma}^{}\nonumber\\
&+V_{31}(\vec{p}_{1},...,\vec{p}_{4};\sigma,\sigma')\alpha_{\vec{p}_{3},\sigma}^{\dag}\alpha_{\vec{p}_{1},\sigma}^{}\alpha_{-\vec{p}_{4},-\sigma'}^{}\alpha_{\vec{p}_{2},\sigma'}^{}
\nonumber\\&+V_{22}^{(1)}(\vec{p}_{1},...,\vec{p}_{4};\sigma,\sigma')\alpha_{\vec{p}_{3},\sigma}^{\dag}\alpha_{\vec{p}_{4},\sigma'}^{\dag}\alpha_{\vec{p}_{2},\sigma'}^{}\alpha_{\vec{p}_{1},\sigma}^{}
\nonumber\\&+V_{22}^{(2)}(\vec{p}_{1},...,\vec{p}_{4};\sigma,\sigma')\alpha_{\vec{p}_{3},\sigma}^{\dag}\alpha_{\vec{p}_{4},-\sigma'}^{\dag}\alpha_{\vec{p}_{2},\sigma'}^{}\alpha_{\vec{p_{1}},-\sigma}^{},
\nonumber\\& \qquad\qquad\qquad\qquad\qquad\qquad\qquad\qquad\qquad\quad\text{(A1)}
\end{align*}
when Eq. \!\!\!(A1) is substituted in Eq. \!\!\!\eqref{Transition probability}, only the term $\alpha_{\vec{p}_{3},\sigma}^{\dag}\alpha_{\vec{p}_{4},-\sigma'}^{\dag}\alpha_{-\vec{p}_{2},\sigma'}^{\dag}\alpha_{\vec{p}_{1}
,\sigma}$ will be entered the calculations. By taking all the permutations of the creation operators into account we get
\begin{align*}
&W_{13}(\vec{p}_{1},\vec{p}_{2},\vec{p}_{3},\vec{p}_{4};\sigma,\sigma')=\frac{1}{4}|(u_{\vec{p}_{4}}u_{\vec{p}_{3}}u_{\vec{p}_{1}}v_{\vec{p}_{2}}+
v_{\vec{p}_{4}}v_{\vec{p}_{3}}v_{\vec{p}_{1}}u_{\vec{p}_{2}})\nonumber\\
&\times[\sigma'(V_{\vec{p}_{3}-\vec{p}_{1}}+V_{\vec{p}_{1}-\vec{p}_{3}})-\sigma'\delta_{\sigma,\sigma'}(V_{\vec{p}_{4}-\vec{p}_{1}}+V_{\vec{p}_{1}-\vec{p}_{4}})]\nonumber\\
&+(u_{\vec{p}_{3}}u_{\vec{p}_{2}}u_{\vec{p}_{1}}v_{\vec{p}_{4}}+v_{\vec{p}_{3}}v_{\vec{p}_{2}}v_{\vec{p}_{1}}u_{\vec{p}_{4}})\nonumber\\
&\times[-\sigma'(V_{\vec{p}_{3}-\vec{p}_{1}}+V_{\vec{p}_{1}-\vec{p}_{3}})+\sigma'\delta_{\sigma,-\sigma'}(V_{-\vec{p}_{1}-\vec{p}_{2}}+V_{\vec{p}_{1}+\vec{p}_{2}})]\nonumber\\
&+(u_{\vec{p}_{4}}u_{\vec{p}_{2}}u_{\vec{p}_{1}}v_{\vec{p}_{3}}+v_{\vec{p}_{4}}v_{\vec{p}_{2}}v_{\vec{p}_{1}}u_{\vec{p}_{3}})\nonumber\\
&\times[\sigma'\delta_{\sigma,\sigma'}(V_{\vec{p}_{4}-\vec{p}_{1}}+V_{\vec{p}_{1}-\vec{p}_{4}})\nonumber\\
&-\sigma'\delta_{\sigma,-\sigma'}(V_{-\vec{p}_{1}-\vec{p}_{2}}+V_{\vec{p}_{1}+\vec{p}_{2}})]|^{2},\text{\,\qquad\qquad\qquad\qquad(A2)}
\end{align*}
in general, as is seen from Eq. \!\!(A2) $W_{13}$ is, a function of all the momenta through potentials, spins and temperature. Since the method of calculations is the same for the other processes, we shall write only the results of them.
\begin{align*}
W&_{31}(\sigma,\sigma')\nonumber\\
&=\frac{1}{4}|(v_{\vec{p}_{3}}v_{\vec{p}_{4}}v_{\vec{p}_{1}}u_{\vec{p}_{2}}+
u_{\vec{p}_{1}}u_{\vec{p}_{3}}u_{\vec{p}_{4}}v_{\vec{p}_{2}})\nonumber\\
&\times[-\sigma'(V_{\vec{p}_{1}-\vec{p}_{3}}+V_{\vec{p}_{3}-\vec{p}_{1}})
+\sigma'\delta_{\sigma,\sigma'}(V_{\vec{p}_{1}+\vec{p}_{2}}+V_{-\vec{p}_{1}-\vec{p}_{2}})]\nonumber\\
&+(v_{\vec{p}_{1}}v_{\vec{p}_{2}}v_{\vec{p}_{4}}u_{\vec{p}_{3}}+u_{\vec{p}_{1}}u_{\vec{p}_{2}}u_{\vec{p}_{4}}v_{\vec{p}_{3}})\nonumber\\
&\times[-\sigma'\delta_{\sigma,-\sigma'}(V_{\vec{p}_{2}-\vec{p}_{3}}+V_{\vec{p}_{3}-\vec{p}_{2}})
+\sigma'(V_{\vec{p}_{1}-\vec{p}_{3}}+V_{\vec{p}_{3}-\vec{p}_{1}})]\nonumber\\
&+(v_{\vec{p}_{2}}v_{\vec{p}_{3}}v_{\vec{p}_{4}}u_{\vec{p}_{1}}+u_{\vec{p}_{2}}u_{\vec{p}_{3}}u_{\vec{p}_{4}}v_{\vec{p}_{1}})\nonumber\\
&\times[-\sigma'\delta_{\sigma,\sigma'}(V_{\vec{p}_{1}+\vec{p}_{2}}+V_{-\vec{p}_{1}-\vec{p}_{2}})\nonumber\\
&+\sigma'\delta_{\sigma,-\sigma'}(V_{\vec{p}_{2}-\vec{p}_{3}}+V_{\vec{p}_{3}-\vec{p}_{2}})]|^{2},\;\,\,\,\qquad\qquad\qquad\quad\text{(A3)}
\end{align*}
and
\begin{align*}
W&_{22}(\sigma,\sigma')\nonumber\\
&=\frac{1}{4}|(V_{\vec{p}_{1}-\vec{p}_{3}}+V_{\vec{p}_{3}-\vec{p}_{1}})[u_{\vec{p}_{1}}u_{\vec{p}_{2}}u_{\vec{p}_{3}}u_{\vec{p}_{4}}\nonumber\\
&+v_{\vec{p}_{1}}v_{\vec{p}_{2}}v_{\vec{p}_{3}}v_{\vec{p}_{4}}-u_{\vec{p}_{2}}u_{\vec{p}_{4}}v_{\vec{p}_{1}}v_{\vec{p}_{3}}-
u_{\vec{p}_{1}}u_{\vec{p}_{3}}v_{\vec{p}_{2}}v_{\vec{p}_{4}}]\nonumber\\
&-(V_{\vec{p}_{2}-\vec{p}_{3}}+V_{\vec{p}_{3}-\vec{p}_{2}})[u_{\vec{p}_{1}}u_{\vec{p}_{2}}u_{\vec{p}_{3}}u_{\vec{p}_{4}}
+v_{\vec{p}_{1}}v_{\vec{p}_{2}}v_{\vec{p}_{3}}v_{\vec{p}_{4}}\nonumber\\
&-u_{\vec{p}_{2}}u_{\vec{p}_{3}}v_{\vec{p}_{1}}v_{\vec{p}_{4}}-u_{\vec{p}_{1}}u_{\vec{p}_{4}}v_{\vec{p}_{2}}v_{\vec{p}_{3}}]\delta_{\sigma,\sigma'}\nonumber\\
&+(V_{\vec{p}_{1}+\vec{p}_{2}}+V_{-\vec{p}_{1}-\vec{p}_{2}})\nonumber\\
&\times[u_{\vec{p}_{1}}u_{\vec{p}_{3}}v_{\vec{p}_{2}}v_{\vec{p}_{4}}+v_{\vec{p}_{2}}v_{\vec{p}_{3}}u_{\vec{p}_{1}}u_{\vec{p}_{4}}
+v_{\vec{p}_{1}}v_{\vec{p}_{4}}u_{\vec{p}_{2}}u_{\vec{p}_{3}}\nonumber\\
&+v_{\vec{p}_{1}}v_{\vec{p}_{3}}u_{\vec{p}_{2}}u_{\vec{p}_{4}}]\delta_{\sigma,-\sigma'}|^{2}.\qquad\qquad\qquad\qquad\qquad\quad\text{(A4)}
\end{align*}
For the Bogoliubov coefficients $u_{\vec{p}}$ and $v_{\vec{p}}$, we insert the standard form
\begin{align*}
\qquad u_{\vec{p}}^{2}=\frac{1}{2}(1+\frac{\varepsilon_{\vec{p}}}{E_{\vec{p}}})\qquad;\qquad v_{\vec{p}}^{2}=\frac{1}{2}(1-\frac{\varepsilon_{\vec{p}}}{E_{\vec{p}}}),\qquad\text{(A5)}
\end{align*}
at low temperature limit, because more quasiparticles locate in the nodes of gap, thus Bogoliubov coefficients may be written as $u_{\vec{p}}\simeq1$ and $v_{\vec{p}}\simeq0$. By using these approximations, transition probabilities can be calculated as
\begin{align*}
\qquad\qquad \quad &W_{13}(\uparrow\downarrow)\simeq0\quad;\quad W_{13}(\upuparrows)\simeq0\nonumber\\
&W_{31}(\upuparrows)\simeq0\quad;\quad W_{31}(\uparrow\downarrow)\simeq0,\quad\qquad\quad\text{(A6)}
\end{align*}
\begin{align*}
\qquad\qquad W_{22}(\uparrow\downarrow)=\frac{1}{4}|(V_{\vec{p}_{1}-\vec{p}_{3}}+V_{\vec{p}_{3}-\vec{p}_{1}})|^{2},\qquad\qquad\text{(A7)}
\end{align*}
\begin{align*}
W_{22}(\upuparrows)=\frac{1}{4}|(V_{\vec{p}_{1}-\vec{p}_{3}}+V_{\vec{p}_{3}-\vec{p}_{1}})-(V_{\vec{p}_{3}-\vec{p}_{2}}+V_{\vec{p}_{2}-\vec{p}_{3}})|^{2}.
\\ \qquad\qquad\qquad\qquad\quad\qquad\qquad\qquad\qquad\qquad\qquad\qquad\text{(A8)}
\end{align*}

\end{document}